
\baselineskip 12 pt
\magnification=1200
\tolerance 10000
\def\S{\Sigma}
\def\s{\sigma}
\def\ep{\epsilon}
\def\b{\beta}
\def\dpfour{  {d^4p\over (2\pi)^4} \>}
\def\dqfour{  {d^4q\over (2\pi)^4} \>}
\def\dkfour{  {d^4k\over (2\pi)^4} \>}
\def\dpthree{  {d^3p\over (2\pi)^3} \>}
\def\dq0{ {dq_0\over 2\pi} }
\def\p{\vec p}
\def\q{\vec q}
\def\ipxy{i\vec p . (\vec x - \vec y) }
\def\op{ \omega_p }
\def\l{ \lambda }
\def\om{  \omega }

\hfill MIT-CTP-2374

\hfill 10/94

\vskip 3 truecm
\centerline{\bf THERMALIZATION AND PINCH SINGULARITIES IN}
\bigskip
\centerline {{\bf NON-EQUILIBRIUM QUANTUM FIELD THEORY}
\footnote{*}{This work is supported in part by funds
provided by the U.S. Department of Energy (D.O.E.) under
cooperative research
agreement DE-FC02-94ER40818.}}
\vskip 2 truecm
\centerline{P. F. Bedaque }
\bigskip
\centerline{\it Center for Theoretical Physics}
\centerline{\it Laboratory for Nuclear Science}
\centerline{\it  and Department of Physics}
\centerline{\it Massachusetts Institute of Technology}
\centerline{\it Cambridge, Massachusetts 02139}
\vskip 2.5 truecm
\centerline{ABSTRACT}
\vskip 1 truecm

I argue that, within the Closed-Time-Path formalism,
pinch singularities do not
appear in truly out of equilibrium situations.

\vfill
\eject

There are two main diagrammatic methods for the study of finite temperature
field theory, the so called imaginary formalism (ITF) [1,2] and the real time
formalism (RTF) [3,4,5,6,7]. The RTF has many versions (Thermofield Dynamics
[7],
Closed-Time-Path Formalism (CTPF) [3,4,6], Influence Functional [8]) that are
equivalent to each order, at least perturbatively. The main advantage of the
RTF over the most popular ITF is that it can be easily extended to deal
with systems that are out of thermal equilibrium. The perturbative series
generated by the RTF can be organised in Feynman diagrams very much like the
usual zero temperature field theory, the only difference being that the
propagators acquire a $2 \times 2$ matrix structure having by entries Green's
functions satisfying different boundary conditions. This immediately raises the
question of the existence of pinch singularities: since propagators with
different boundary conditions are used, what prevents ill defined terms like

$$\int d^4p\> {1\over p^2-m^2+i\ep}{1\over p^2-m^2-i\ep}\eqno(1)$$

\noindent
from appearing on a diagram ?
The existence of these singularities would be a disaster to perturbative
calculations done in the CTPF framework.
Actually, an early attempt of studying finite
temperature field theory in real time [9] (still without the matrix
propagators)
failed beyond one loop because of singularities of this kind and led to the
rediscovery of the CTPF. In equilibrium, it is known that, after the inclusion
of the matrix propagators, these singularities cancel. However, it was recently
pointed out [10,11] that this cancelation works due to the precise form of the
equilibrium particle distribution function

$$n(\om )={1\over e^{\b\om} -1}\ .\eqno(2)$$

\noindent
Any other distribution function and, in particular, a time dependent one, would
lead to uncanceled pinch singularities in the perturbative expansion.
Does that mean that perturbation theory breaks down in non-equilibrium
processes ?

In this letter I shall argue that when physically meaningful out of
equilibrium situations are analysed, pinch singularities do not appear
and demonstrate this on a example.

 To make the discussion more concrete, let us consider a model with
two scalar fields $\phi$ and $\s$ with masses $m$ and $M$  that are
free for times $t<0$ and at thermal equilibrium at temperatures $T$ and $T'$
respectively.
At $t=0$ an interaction of the form $-\l\phi\s^n$ is turned on.
Through their mutual
interaction, the fields will then thermalize.
This process, however, takes time.
At any finite time after the interaction starts, the Green's functions will
reflect the non-equilibrium nature of the problem. For instance, since time
translation invariance is broken, the propagator $D(x,y)$
will depend on the time
coordinates $x^0$ and $y^0$ separately and not only on
the difference $x^0-y^0$.
This and similar models have been analysed in the
literature before [12,13,14].
The Lagrangian is

$$L={1\over 2} \partial_\mu \phi \partial^\mu \phi - {m^2\over 2} \phi^2 +
       {1\over 2} \partial_\mu \s \partial^\mu \s - {m^2\over 2} \s^2-
       \theta (t) \l \phi\s^n\ ,\eqno(3)$$

\noindent
where $\theta (t)$ is the step function.
The Lagrangian (3) prepares the system in an initial condition at $t=0$ that is
not in equilibirum, that is, the density matrix describing it at $t=0$ does not
commute with the Hamiltonian for $t>0$.

We want to find the propagator of the $\phi$ field up to order $\l^2$.
In the CTPF, the free Lagrangian around which the
perturbation expansion is built is taken as the quadratic part of the
Lagrangian at minus infinity.  Three, four,...etc.,  particle correlations
(inexistent in our model) present in the
initial condition as well as the interactions
present at $t>0$
are treated as perturbations.
The bare $\phi$ field propagator is obtained from the
quadratic part of (3) [6,15]:

$$\eqalign{iD^0(x,y)=&\left(\matrix{
                          <T\phi(x)\phi(y)> & <\phi(x)\phi(y)>  \cr
                          <\phi(y)\phi(x)>  & <T^*\phi(x)\phi(y)> \cr
                         }\right) \cr
                   =&\left(\matrix{
                      {i\over p^2-m^2+i\ep} + 2\pi n(|p_0|)\delta(p^2-m^2)&
                       2\pi(\theta(-p_0)+n(|p_0|))\delta(p^2-m^2) \cr
                       2\pi(\theta(p_0)+n(|p_0|))\delta(p^2-m^2) &
                       -{i\over p^2-m^2-i\ep} + 2\pi n(|p_0|)\delta(p^2-m^2)
\cr
                         }\right)\ , \cr}\eqno(4)$$

\noindent
where ($T^*$) $T$ is the (anti-)time ordering operator and $n(p_0)$ is the
equilibrium bosonic distribution function at temperature $T=1/\b$

$$n(p_0)={1\over e^{\b p_0}-1} \ .\eqno(5)$$

\noindent
In CTPF, there are two kinds of vertices, one connecting the
$1-1$ components of the
propagators ($-i\l \theta(t) \phi\s^n$) and another with opposite sign
connecting the $2-2$ components  ($i\l \theta(t) \phi\s^n$).
The connection between the bare propagator $D^0$, the full propagator $D$ and
the self-energy is

$$D=( D^{0\ -1} + \S)^{-1}\ . \eqno(6)$$

\noindent
We are using a notation where $D(x,y)$ is seen as an infinite matrix.
For example, $ (DD')(x,y)$ stands for $\int dz D(x,z)D'(z,y)$.
It is convenient to change
basis in  (4) and (6) performing a similarity transformation by the matrix

$$V={1\over \sqrt{2}}\left(\matrix{
                                   1 & 1 \cr
                                  -1 & 1 \cr
                                 }\right)\ , \eqno(7) $$

\noindent
after which equation (6) reads

$$\left(\matrix{
                0  & D_A \cr
                D_R& D_c \cr
               }\right)
      =\left[\left(\matrix{
                           0  & D_A \cr
                           D_R& D_c \cr
                           }\right)^{-1}
      +\left(\matrix{
                      \S_c & \S_R \cr
                      \S_A & 0
                    }\right)\right]^{-1}\ ,\eqno(8)$$

\noindent
where $D_A$,$D_R$ and $D_c$ are the retarded, advanced and correlated Green's
functions

$$\eqalign{iD_R(x,y)=&\theta(x^0-y^0)<[\phi(x),\phi(y)]>\ ,\cr
           iD_A(x,y)=&\theta(y^0-x^0)<[\phi(x),\phi(y)]>\ ,\cr
           iD_c(x,y)=&<[\phi(x),\phi(y)]_+>\ ,\cr}\eqno(9)$$

\noindent
and

$$\eqalign{\S_R=&\S_{11}+\S_{12}\ ,\cr
           \S_A=&-\S_{12}-\S_{22}\ ,\cr
           \S_c=&\S_{11}+\S_{22}\ .\cr}\eqno(10)$$

\noindent
Equation (8) implies

$$\eqalignno{
D_R=&(D_R^{0\ -1} + \S_R)^{-1}\sim D_R^0-D_R^0\S_R D_R^0+O (\S^2)\
,&(11a)\cr
D_A=&(D_A^{0\ -1} + \S_A)^{-1}\sim D_A^0-D_A^0\S_A D_A^0+O (\S^2)\
,&(11b)\cr
D_c=&D_R D_R^{0\ -1} D_c^0 D_A^{0\ -1} D_A - D_R\S_cD_A&\cr
    \sim & D_c^0-
   D_R^0\S_RD_c^0 - D_C^0\S_A D_A^0 - D_R^0\S_c D_A^0 + O (\S^2)\
. &(11c)\cr}$$

In our model, $\S$ will be given in lowest order by a graph with $n$
$\s$-lines joined at their ends
(for $n\le 3$ there are also graphs with tadpoles).
Since the interaction exists only for positive times, $\S$ will be given by

$$\S(x,y)=\theta(x^0) \bar \S (x-y) \theta(y^0)\ ,\eqno(12)$$

\noindent
where $\bar\S$ is the self-energy for the system in equilibrium, that means,
the self-energy for a system with constant coupling $\l$.
The specific form of $\bar\S$ will not be needed in our argument.
Let us consider
the perturbative expansion of $D_c$ (11c). Some insight can be gained  looking
at (11c) in  position space.
For example, one of the terms  in (11c) is

$$(D_R^0\S_cD_A^0)(x,y) = \int d^4z\ d^4z' \ \
                  D_R^0(x-z) \theta(z^0)\bar\S_c(z-z')D_A^0(z'-y)
\theta({z^0}')
       \ .\eqno(13)$$

\noindent
Since $D_R (D_A)$ vanishes for $z^0>x^0 (y^0<{z^0}')$, the integration on
$z^0 ({z^0}')$ is bounded from above by $x^0 (y^0)$ and from below by $0$.
A similar thing happens with the other two terms in (11c),
and the range of the $z^0, {z^0}'$ integration
is limited there too. We have then that $D_c$ is given in first order by an
integral of well behaved functions on a bounded domain: there are no room for
divergencies (of course, the integration on $\vec z, {\vec z}'$
may produce the usual ultraviolet divergencies, unrelated to the pinch
singularities). Note that this would
not be the case had the interaction existed
for all times. The unlimited range of integration would have produced
divergencies that in momentum space show up as pinch singularities. But
fields that have interacted since $t=-\infty$ should have attained equilibrium
by any finite time. One either has fields that started interacting at some
finite time (and then the range of integration is limited and no singularities
arise) or fields that were in contact since the beginning of times ( in which
case they should be thermalized at any finite time considered and the pinch
singularities cancel as shown in [10]). The situation with fields at two
different temperatures in contact with each other for an in finite amount
of time is clearly unphysical.
The perturbative expansion of $D_R$ and $D_A$ also always involve integrations
on a limited domain and never contain pinch singularities.

These remarks will be  clearer in a more explicit calculation in momentum
space. In particular it will be clear that pinch singularities are actually
related to the unboundness of the domain of integration.

Up to leading order,
$D_c$ (11c) involves products of three functions of the form

$$\eqalign{ (D^0_i\S_j D^0_k)(x,y)=&\int d^4z\ d^4z' D^0_i (x-z) \theta(z^0)
                                 \bar\S_j (z-z') \theta({z^0}') D^0_k (z'-y)\cr
                              =&\int dz\ dz' \dpfour\dkfour\dqfour
                                            \theta(z^0)\theta({z^0}')
                              e^{-ip(x-z)-ik(z-z')-iq(z'-y)}\cr
                     &\phantom{ \int dz\ dz' \dpfour\dkfour\dqfour}\times
                            D^0_i (p)\bar\S_j (k) D^0_k (q)\cr}\ ,\eqno(14)$$

\noindent
where $i,j,k=A,R,c$. Equation (14) can be evaluated with the help of

$$\int^\infty_0 dz \ e^{i\alpha z}=\lim_{\eta\rightarrow 0^+}
                                   \ \  {i\over \alpha+i\eta}\eqno(15)$$

\noindent
giving

$$\eqalign{(D^0_i\S_j D^0_k)(x,y)=& \lim_{\eta\rightarrow 0^+}\int\dpfour
                         {dq_0\over 2\pi}
                    e^{-ip_0x_0+iq_0y_0+\ipxy} {1\over p_0-q_0+2i\eta}\cr
                &\phantom{ \lim_{\eta\rightarrow 0^+}\int\dpfour
                                  {dq_0\over 2\pi} }
                \times\     D^0_i (p_0,\p)\bar\S_j (q_0,\p) D^0_k (q_0,\q)\ .
                   \cr} \eqno(16)$$

\noindent
The limited range of integration on $z^0,{z^0}'$ produces
the denominator in (16)
instead of the usual $\delta(p_0-q_0)$
and the energy is not conserved in the
vertices. The physical origin of this is clear, at the instant $t=0$ the system
suddenly acquires some " potential " energy it did not have before the
interaction was turned on. It may seem that the substitution of
$\delta(p_0-q_0)$ by $(p_0-q_0+2i\eta)^{-1}$ is not enough to get rid of the
pinch singularities since

$${i\over p_0-q_0+2i\eta}={\cal P} \left( {i\over p_0-q_0} \right)+
                          \pi \delta(p_0-q_0)\ ,\eqno(17)$$

\noindent
and the term with the $\delta$ function, when used in (16), produces a term
proportional to $D^0_i (p_0,\p)\bar\S_j (p_0,\p) D^0_k (p_0,\q)$, that contains
pinch singularities.
An explicit calculation will show that the term coming from the
principal value
in (17) is also singular in a way that the whole expression (16) is well
defined.

First let us calculate the term in (11c) giving by (16) with $i=R, j=c$ and
$k=A$ (we are assuming from on that $x^0>y^0$,
the other cases can be handled
similarly):

$$\eqalign{(D^0_R\S_c& D^0_A)(x,y)=\cr
       = & \lim_{\ep\rightarrow 0^+}\lim_{\eta\rightarrow 0^+}
     \int    \dpfour \dq0  e^{-ip_0x_0+iq_0y_0+\ipxy} {1\over
p_0-q_0+2i\eta}\cr
 &\phantom{ \lim_{\ep\rightarrow 0^+}\lim_{\eta\rightarrow 0^+}
     \int    \dpfour \dq0    }\times
 {i\over (p_0+i\ep)^2-\op^2}{i\over(q_0-i\ep)^2-\op^2}\bar\S_c(q_0,\p)\cr
         =&\lim_{\ep\rightarrow 0^+}\int\dpfour e^{\ipxy}
                 \Biggl[ e^{-ip_0(x_0-y_0)}
      {i\over (p_0+i\ep)^2-\op^2}{i\over(p_0-i\ep)^2-\op^2}\bar\S_c(p_0,\p)\cr
       \ -&  e^{-ip_0x_0+i(\op+i\ep)y_0} {i\over p_0-\op-i\ep}
          {i\over (p_0+i\ep)^2-\op^2}{1\over 2(\op+i\ep)}
                          \bar\S_c(\op+i\ep,\p)\cr
       \ -& e^{-ip_0x_0+i(-\op+i\ep)y_0} {i\over p_0+\op-i\ep}
          {i\over (p_0+i\ep)^2-\op^2}{1\over 2(-\op+i\ep)}
                              \bar\S_c(-\op+i\ep,\p)\cr
       \ +& e^{-ip_0x_0+i\ep_p y_0}{i\over p_0-\ep_p}{i\over
(p_0+i\ep)^2-\op^2}
          {i\over \ep_p^2 - \op^2}\bar\S^r_c(\ep_p,\p) \Biggr]\ ,
                             \cr}\eqno(18)$$

\noindent
where $\op=\sqrt{\p^2+m^2}$ and $\bar\S^r_c(\ep_p,\p)$ is the residue of
$\bar\S_c$ at its pole $\ep_p$ (in the upper half plane).
The first term in (18) is all we would have in the case of time independent
coupling. There is a pinch singularity there. However, the remaining terms are
singular too, and the singularities cancel. In fact, the integration on $p_0$
gives

$$\eqalign{(D^0_R\S_c D^0_A)(x,y)=\int \dpthree e^{\ipxy}
   \Biggl[& -e^{-i\op(x_0-y_0)-\ep(x_0-y_0)}{1\over 4\ep}{1\over \op-i\ep}
                                                     \bar\S_c(\op-i\ep,\p)\cr
   &-e^{i\op(x_0-y_0)-\ep(x_0-y_0)}{1\over 4\ep}{1\over \op+i\ep}
                                                  \bar\S_c(-\op-i\ep,\p)\cr
   &+e^{-i\op(x_0-y_0)-\ep(x_0+y_0)}{1\over 4\ep}{1\over \op}
                                                 \bar\S_c(\op+i\ep,\p)\cr
   &+e^{i\op(x_0+y_0)}{i\over 8}{1\over \op^3}    \bar\S_c(-\op,\p)\cr}$$

$$\eqalign{\phantom{  (D^0_R\S_c D^0_A)(x,y)=\int \dpthree e^{\ipxy}
   \Biggl[  }
   &+e^{-i\op(x_0+y_0)}{i\over 8}{1\over \op^3}   \bar\S_c(-\op,\p)\cr
   &+e^{i\op(x_0-y_0)-\ep(x_0+y_0)}{1\over 4\ep}{1\over \op}
                                                  \bar\S_c(-\op+i\ep,\p)\cr
   &+e^{-i\op x_0+i\ep_p y_0} {i\over \op-\ep_p}{1\over \ep_p^2-\op^2}
        {1\over 2\op}\bar\S^r_c(\ep_p,\p)\cr
   &+e^{i\op x_0+i\ep_p y_0} {i\over \op+\ep_p}{1\over \ep_p^2-\op^2}
        {1\over 2\op}\bar\S^r_c(\ep_p,\p)\cr
   &+e^{-i\ep_p(x_0-y_0)}{i\over (\ep_p^2-\op^2)^2}\bar\S^r_c(\ep_p,\p)
           \Biggr]\cr}\ .\eqno(19)$$

\noindent
In (19), the first two lines correspond to the fist line in (18). Thus, in
accord to the general argument above, the pinch singularities cancel in this
expression. Note that the terms that depend on $x_0$ and $y_0$ separately,
and not only on the difference $x_0-y_0$, come from the terms in (18) that
would not be present if the coupling were constant. Those terms are the ones
that break time translation invariance and correspond to truly non-equilibrium
effects.

The same kind of cancelation occurs in the other two terms on (11c).
We just quote the results of the integrations here:

$$\eqalign{(D^0_R\S_R D^0_c)(x,y)=\int &\dpthree  e^{\ipxy} \Biggl[
    e^{-i(\op-i\ep)(x_0-y_0)}{1\over 4\op\ep} f(\op-i\ep)\cr
    &-e^{i(\op+i\ep)(x_0-y_0)}{1\over 4\op\ep} f(-\op-i\ep)\cr
    &+{d\over dq_0}{i\over (q_0+\op)^2}\Bigl(e^{-iq_0(x_0-y_0)}f(q_0)
                                           +e^{iq_0(x_0-y_0)}f(-q_0)\Bigr)
                                              \Biggl\vert_{q_0=\op}\cr
    &-e^{-i\op(x_0-y_0)-\ep(x_0+y_0)}{1\over 4\op\ep} f(\op+i\ep)\cr
    &+e^{-i\op(x_0+y_0)}{i\over 8\op^3} f(\op)\cr
    &+e^{i\op(x_0-y_0)-\ep(x_0+y_0)}{1\over 4\op\ep} f(-\op+i\ep)\cr
    &+e^{i\op(x_0+y_0)}{i\over 8\op^3} f(\op)\Biggr]\ ,\cr}\eqno(20)$$

\noindent
and
 $$(D^0_c\S_A D^0_A)(x,y)=(D^0_R\S_R D^0_c)(y,x)\ ,\eqno(21)$$

\noindent
where

$$f(q_0)=(1+2n(q_0))\bar\S_R(q_0,\p)\ .\eqno(22)$$

\noindent
Here too, the new terms generated by the time dependence of the coupling
contain a part dependent only on the difference $x_0-y_0$ that cancels the
singularity of the " equilibrium " term and the result is well defined.
Note that, at no point we used the specific form of $\S$,
so this cancelation seems to be
a general feature.
In particular it is independent of the specific distribution function of the
$\s$-particles and its initial temperature.
For $n=1$ our model is trivially solved and the result,
expanded to order $\l^2$ agrees with (19), (20) and (21).
%

I have argued that the perturbation series is well defined and free of pinch
singularities. That does not mean though that perturbation theory is adequate
for all non-equilibrium phenomena. Looking at (19) and (20) we see that $D_c$
calculated up to order $\l^2$ contain terms proportional to $x_0$ and $y_0$.
Higher order calculations of $D_{R(A)},D_c$ would contain higher powers
of $x^0,y^0$. This is the way perturbation theory has of approximating the true
large time behavior of the propagator (dumped exponential) we expect on
physical grounds. Any finite order in perturbation theory will do a poor job in
approximating this dumped behavior at large times. This is analogous to the
fact that the series $1-\gamma t + \gamma^2 t^2/2 -\dots$ is not a good
approximation to $e^{-\gamma t}$ for $t>>\gamma^{-1}$. A solution
for this problem is to use full propagators
$D_{R(A)}=( {D^0_{R(A)} }^{-1} + \S_{R(A)})^{-1}$, with $\S_{R(A)}$
calculated in perturbatively,  instead of the bare ones.
This resums these powers of $x^0,y^0$ into an dumped exponential we expect
for stable systems.
One may feel tempted though, to lift the restriction imposed by $\theta (z^0),
\theta ({z^0}')$  in , for instance, equation (13), since the contribution
coming from $z^0,{z^0}'< 0$ is small for large $x^0,y^0$ due to the
dumped exponential behavior of $D_{R(A)}$.
This is just a consequence of the fact that for large times the system looses
memory of the initial condition at $t=0$ (the system thermalizes). But
in doing this approximation, namely, integrating over $z^0,{z^0}'$ from
$-\infty $ to $\infty$, one throws away all pieces that break time translation
invariance and the result is just the thermalized (equilibrium) Green's
functions. For short times (shorter than the thermalization time), where the
essentially non-equilibrium behavior occurs, perturbation
theory works finely,  but
it is important to keep the restriction $z^0,{z^0}>0$. This way the
non-equilibrium physics (breaking of time translation invariance and the
dependence on initial conditions) is preserved and pinch singularities do not
appear.

The author would like to thank A.Das for comments.
\hfill
\eject
{\bf REFERENCES}
\bigskip
\noindent
[1] J.Negele and H.Orland {\it Quantum Many-Particle Systems}, Addison-Wesley
(1988)
\medskip
\noindent
[2] J.Kapusta, {\it Finite Temperature Field Theory},  Cambridge Press,1989.
\medskip
\noindent
[3] J.Schwinger, {\it J.Math.Phys.} 2 (1961) 407.

\medskip
\noindent
[4] L.V.Keldysh, {\it Sov.Phys.} 20 (1964) 1018.

\medskip
\noindent
[5] P.M.Bakshi and K.T.Mahanthappa, {\it J.Math.Phys}4 (1963) 1, 12.

\medskip
\noindent
[6] K.C.Chou et al. {\it Phys.Rep.} 118 (1985) 1.

\medskip
\noindent
[7] H.Umezawa, H.Matsumoto and M.Tachiki, Thermo Field Dynamics and Condensed
States (North-Holland, Amsterdam, 1982).

\medskip
\noindent
[8] R.P.Feynman and  F.C.Vernon, {\it Ann.Phys.}, 24 (1963) 118.

\medskip
\noindent
[9] L.Dolan and R.Jackiw, {\it Phys.Rev.}D9, 3320 (1974).

\medskip
\noindent
[10] T.Altherr and D.Seibert, {\it Phys.Lett.}B, 333 (1994) 149.

\medskip
\noindent
[11] H.A.Weldon, {\it Phys.Rev.}D45 (1992) 352.

\medskip
\noindent
[12] E.Calzetta, {\it Ann.Phys.}(N.Y.), 190, (1989) 32.

\medskip
\noindent
[13] D.Boyanovsky, D.-S.Lee and A.Singh,{\it Phys.Rev.}D48, 800 (1993).

\medskip
\noindent
[14] P.F.Bedaque and A.Das, {\it Mod.Phys.Lett.}A8, 3151, (1993).

\medskip
\noindent
[15] A.Niemi and G.Semenoff, {\it Ann.Phys.} 152, 105 (1984).
\end